\documentclass[twocolumn,aps,prl,superscriptaddress,amsfonts,amssymb,amsmath]{revtex4}
\usepackage{graphicx}
\begin{document}
\title{\bf Partial entropy in  finite-temperature phase transitions}
\author{Junpeng Cao}
\affiliation{Beijing National Laboratory for Condensed Matter
Physics, Institute of Physics, Chinese Academy of Sciences,
Beijing 100080, PR China}
\author{Xiaoling Cui}
\affiliation{Beijing National Laboratory for Condensed Matter
Physics, Institute of Physics, Chinese Academy of Sciences,
Beijing 100080, PR China}
\author{Qi Zhang}
\affiliation{Beijing National Laboratory for Condensed Matter
Physics, Institute of Physics, Chinese Academy of Sciences,
Beijing 100080, PR China}
\author{Wengang Lu}
\affiliation{Beijing National Laboratory for Condensed Matter
Physics, Institute of Physics, Chinese Academy of Sciences,
Beijing 100080, PR China}
\author{Qian Niu}
\affiliation{Department of Physics, The University of Texas,
Austin, Texas 78712-1081 USA} \affiliation{International Center
for Quantum Structures, Chinese Academy of Sciences, Beijing
100080, PR China}
\author{Yupeng Wang$^*$}
\affiliation{Beijing National Laboratory for Condensed Matter
Physics, Institute of Physics, Chinese Academy of Sciences,
Beijing 100080, PR China} \affiliation{International Center for
Quantum Structures, Chinese Academy of Sciences, Beijing 100080,
PR China}

\begin{abstract}
It is shown that the von Neumann entropy, a measure of quantum
entanglement, does have its classical counterpart in thermodynamic
systems, which we call partial entropy. Close to the critical
temperature the partial entropy shows perfect finite-size scaling
behavior even for quite small system sizes. This provides a
powerful tool to quantify finite-temperature phase transitions as
demonstrated on the classical Ising model on a square lattice and
the ferromagnetic Heisenberg model on a cubic lattice.
\end{abstract}


\maketitle

\par
Recently, it is found  that quantum phase transitions can be
quantified from the inflexions of the quantum entanglement
measures
\cite{qpt1,qpt2,qpt3,qpt4,qpt5,qpt6,qpt7,qpt8,qpt9,qpt10}. Close
to the quantum critical point, the von Neumann entropy, a measure
of the quantum entanglement, shows finite-size scaling behavior.
This method is quite powerful and straightforward for finding
quantum phase transitions in quantum models because one needs
neither  a pre-assumed order parameter nor a considerable large
system size. A natural question arises: Is it possible to
generalize this method to quantify finite-temperature phase
transitions in thermodynamic systems?

There are several established methods to investigate the
finite-temperature phase transitions, such as exact solutions,
mean field approach, series expansion, renormalization group
analysis, and numerical evaluation of the partition function or
correlation functions. However, each known method has its
shortcoming. Only few models are exactly soluble. The mean field
approach needs a pre-assumed order parameter and may not be
reliable. All the numerical methods and the renormalization group
analysis need to study large system sizes with complicated
computation processes. Therefore, it is highly desirable to find
an efficient and general method to characterize finite-temperature
phase transitions in both quantum and classical systems.

In this Letter, we point out that the partial entropy, a
counterpart of the von Neumann entropy for thermodynamic systems,
captures the common feature of all phase transitions, i.e., the
information of critical fluctuations.  With two models (one
classical and one quantum), we show that close to the critical
temperature the derivative of the partial entropy shows perfect
finite-size scaling behavior even for quite small system sizes.
The critical temperature and critical exponents can be determined
by the inflexion or the scaling law of the partial entropy. This
provides a powerful tool to quantify finite-temperature phase
transitions in a variety of interesting models in condensed matter
physics.

In the study of quantum phase transitions, one is concerned with
the ground state properties as described by the density matrix
$|\Psi_0\rangle \langle \Psi_0|$. Crucial information on quantum
correlation or entanglement between a subsystem $p$ and the rest
$\bar p$ exists in the reduced density matrix
$\rho_p(\alpha)\equiv tr_{\bar p}|\Psi_0\rangle \langle \Psi_0|$,
as is captured in the von Neumann entropy,
$E_v=-tr\rho_p\ln\rho_p$. It has been shown that singular behavior
in the von Neuman entropy occurs as a function of control
parameters as the system goes through a quantum phase transition
\cite{qpt0}.

The concept of von Neumann entropy can be straightforwardly
generalized to thermodynamic systems at finite temperatures $T$.
The density matrix now reads $\rho(T)=\exp{(- H/T)}/Z$, where $H$
is the Hamiltonian, and $Z$ is the partition function. One can
similarly define a reduced thermal density matrix
$\rho_p(T)=tr_{\bar p}\rho(T)$, and consider the partial entropy $
S_p(T)\equiv-tr\rho_p(T)\ln\rho_p(T)=-\sum_{n=1}^{D_H} p_n\ln
p_n$, where $D_H$ is the dimension of the Hilbert space of the
subsystem $p$, and $p_n$ are the eigenvalues of $\rho_p(T)$. For a
classical system, the trace operation is replaced by summing over
the classical states. The partial entropy is determined by the
probability distribution of the subsystem. It also measures the
quantum and classical correlation between the subsystem and the
rest of the system. As is shown below, the partial entropy in fact
captures the main feature of the critical fluctuation and
therefore shows singular behavior close to the critical
temperature. Its inflexion gives the information of the critical
point.

As our first example, we study the two-dimensional (2D) Ising
model. It is well known that this system is exactly soluble,
undergoing a second order phase transition \cite{tc} at
$T_c=2/$arcsinh$(1) \approx 2.26919 J/k_B$ and its critical
behaviors have been studied very well. It is therefore an ideal
system to test our method. The Hamiltonian is $H=-\sum_{\langle ij
\rangle } \sigma_{i}^{z}\sigma_{j}^{z}$, where $\sigma_{i}^{z}=\pm
1$ is the spin along $z$-direction on site $i$ and $\langle ij
\rangle$ indicates bonds between nearest neighbor sites.
\begin{figure}[ht]
\begin{center}
\includegraphics[height=6cm,width=8cm]{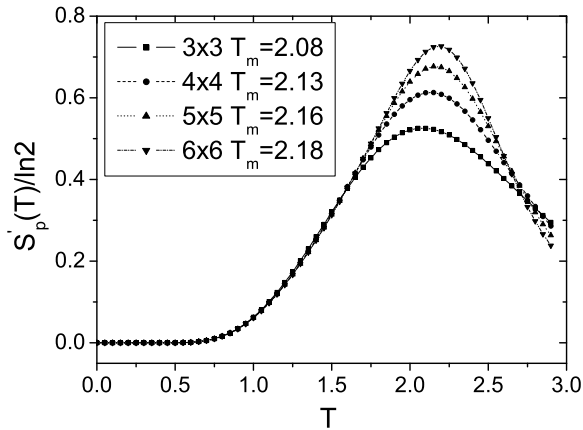}
\end{center}
\caption{The derivative of the partial entropy $S^{\prime}_{p}(T)$
versus temperature for $L=3-6$ are plotted. A maximum appears at
certain temperature $T_m$ for a given $L$. On increasing $L$, the
peak becomes more pronounced and $T_m$ shifts accordingly to the
critical temperature of the 2D Ising model $T_c=2.26919$.}
\label{fig1}
\end{figure}
We consider the $L \times L$ square-lattice case and use the
periodic boundary condition. We focus our attention on the
subsystem of two nearest neighbor sites. The reduced density
matrix is obtained by tracing all the spins except those two.
Since the system is translational invariant, $\rho_p(T)$ is
independent of the choice of the bond and takes the form
$\rho_p(T)=[1+\gamma(T)\sigma_1^z\sigma_2^z]/4$ by a simple
symmetry analysis, where $\gamma(T)$ is a parameter depending on
the temperature and system size $L$. Then we obtain the partial
entropy as
\begin{equation}
S_{p}(T)=2\ln2-\frac{1}{2}\left\{\gamma
\ln\frac{1+\gamma}{1-\gamma}+\ln[1-\gamma^2]\right\}.
\label{entropy}
\end{equation}

\begin{figure}[ht]
\begin{center}
\includegraphics[height=6cm,width=8cm]{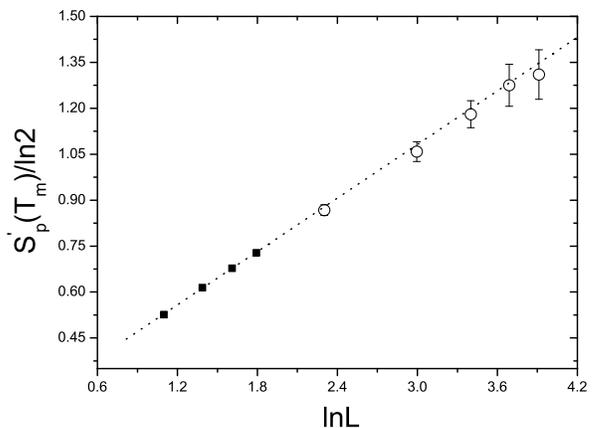}
\end{center}
\caption{The maximum values of the derivative of the partial
entropy $S^{\prime}_{p}(T_m)$ for $L=3-6$ (black dots) depend
linearly on $\ln L$ (dotted line). Monte Carlo results for larger
$L=10,20,30,40,50$ (circles) fall on this straight line,
confirming the logarithmic divergence for a second-order phase
transition.}\label{fig2}
\end{figure}

Thermal fluctuations increase with temperature, so does the
partial entropy. When $T\to 0$, the spins are ordered either up or
down. The parameter $\gamma(0)=1$ and the partial entropy takes
the minimum value of $\ln2$. At extremely high temperatures
$T\to\infty$, all the four states have the same probability to
occur. In this case $\gamma(\infty)=0$ and the partial entropy
takes the maximum value $2\ln2$. The partial entropy increases
monotonically between these two extreme values which are
independent of L.  More interestingly, the derivative of the
partial entropy $S^{\prime}_{p}(T)$ has a maximum, corresponding
to fastest growth of the partial entropy.  The peak at the maximum
sharpens as the system size increases (Fig. \ref{fig1}), and the
maximum value diverges logarithmically with the system size
(Fig.\ref{fig2}).  This singular behavior indicates a critical
point of our system.

\begin{figure}[ht]
\begin{center}
\includegraphics[height=6cm,width=8cm]{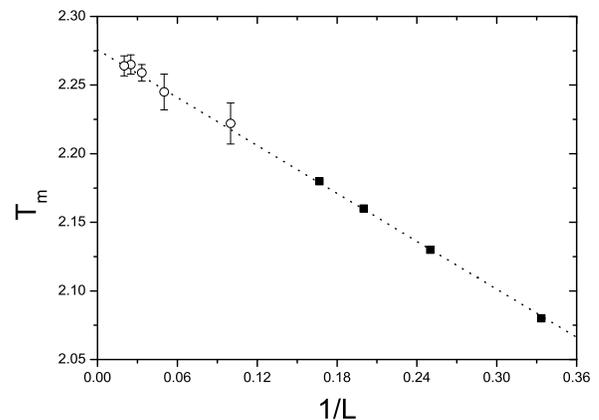}
\end{center}
\caption{The peak temperature $T_m$ versus logarithm of the system
size is depicted to quantify the phase transition temperature
$T_c$. The black dots for $L=3-6$ are from the exact numerical
calculation while the circles for larger $L=10, 20, 30, 40, 50$
are from the Monte Carlo simulation. The dotted line is a fit with
the formula $T_m=2.27802-0.58995 L^{-1}$ from the $L=3-6$ data
only. The $L\to\infty$ limit is very close to the exact value
2.26919. } \label{fig3}
\end{figure}
Exact calculations are performed for system sizes $L=3-6$, and
Monte Carlo simulations are done for large sizes up to $L=50$. The
straight line fit in Fig.\ref{fig2}, $S^{\prime}_{p}(T_m)=0.42034
\ln L + {\text const,}$ is from the first four points, and it is
also followed nicely by the Monte Carlo data. The temperature
$T_m$ at the peak of $ S^{\prime}_p(T)$ shows a size dependence
linear in $1/L$ (Fig.\ref{fig3}). Just from the first four points,
we find the fit $T_m=2.27802-0.58995 L^{-1}$, which turns out to
be also well followed by the Monte Carlo data for larger sizes. A
critical temperature, $T_c=2.27802$, is extracted from the above
fit as the limiting value of $T_m$ at $L\to\infty$. This is very
close to the exact value of $2.26919$. We have thus seen the
effectiveness of the partial entropy method in providing accurate
information on the critical point from fairly small system sizes.

\begin{figure}[ht]
\begin{center}
\includegraphics[height=6cm,width=8cm]{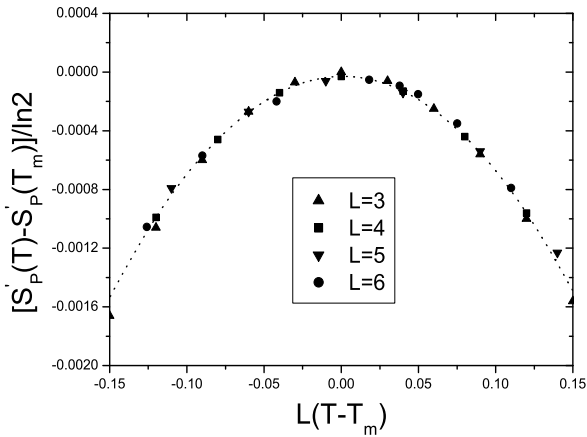}
\end{center}
\caption{The finite size scaling of the partial entropy is
performed. The deviation of the first derivative of the partial
entropy from its maximum is only a function of $L(T-T_m)$. All the
data from $L=3$ to $L=6$ fall on a single curve, indicating a
perfect scaling behavior of the partial entropy around the
critical temperature.} \label{fig4}
\end{figure}
We can also study the scaling behavior of the partial entropy as a
function of both the system size and temperature. We observe that
$S^{\prime}_{p}(T)-S^{\prime}_{p}(T_m) \sim Q[L(T-T_m)]$, where
the scaling function $Q(x) \sim C_{\infty}\ln x$ for large $x$,
and $Q(x) \sim 0$ for very small $x$ \cite{scale}.  We find that
all the data from different system sizes $L$ converge to a single
curve, which is shown in Fig.\ref{fig4}. These results establishes
that finite size scaling is present in the partial entropy.

There is a direct relationship between the partial entropy and
thermodynamic quantities, which explains why singular behavior of
the partial entropy can be used to characterize the critical point
and associated scaling properties.  We observe that the density of
internal energy may be found as $u(T)=L^{-2}tr
H\rho(T)=-2tr_p\rho_p(T)(\sigma_1^z\sigma_2^z)=-2\gamma(T)$.
Therefore, the partial entropy Eq.(\ref{entropy}) can be
determined uniquely by the internal energy density $u$.  Moreover,
the derivative of the partial entropy can be found as
$S^{\prime}_p(T)= Tds(T)\ln[(2-u)/(2+u)]/(4dT)$, where $s(T)$ is
the density of entropy of the whole system and we have used the
thermodynamic relation $du/dT = T ds/dT$. In the neighborhood of
the critical point, the internal energy density $u$ lies somewhere
in the middle of the interval $(-2, 0)$ and varies smoothly with
temperature (with continuous first derivative).  Therefore, the
singular point of $S^{\prime}_p(T)$ coincides with that of the
derivative of the global entropy density $s(T)$.  Therefore, one
can indeed use the partial entropy to quantify thermodynamic phase
transitions.

Moreover, the partial entropy is a quantity more suitable for
finite-size studies than standard thermodynamic quantities, in
that scaling behavior sets in much earlier in the former than in
the latter.  This allows precise determination of the critical
temperature, for example, from relatively small system sizes using
our method.  For contrast, we performed finite-size calculations
for $ds/dT$, finding that its maximum in the temperature
dependence diverges as a polynomial rather than a linear function
of $\ln L$. Also, the peak temperature $T¡¯_m$ of $ds/dT$ is
nonlinear in $1/L$.  The data for $L=3-6$ can be nicely fitted
(with an error less than 0.001) by the function $T¡¯_m=
2.3373-1.50437L^{-2.493}$.  However, its extrapolation (2.3373) to
infinite size is far from the true critical temperature.  One may
take this to mean that the coefficients and the exponent of such a
fit are modulated slowly in $L$ (such as logarithmic). Another
interpretation is that there are subdominant terms in finite-size
scaling which do not decay quickly with the system size.  In any
case, it is very clear that one cannot obtain useful results from
thermodynamic quantities when the system sizes are not very large.
On the other hand, the partial entropy does seem to be free from
such complications.

It remains to be explained why the partial entropy is superior to
the global entropy in finite-size scaling.  At this moment, we do
not have a clear answer, but offer the following observation which
may shed some light to this question.  The basic ingredient of the
partial entropy method is quite similar to that of the
renormalization group transformation. As is well known, in the
later approach, information about the critical fluctuation remains
unchanged, because the critical point is a fixed point of the
renormalization group transformation.  The partial entropy
captures the essence of critical fluctuations of the system.

So far we have been concerned with a classical system, we consider
next a quantum Heisenberg model on a cubic lattice. In this model,
the critical fluctuation is much weaker than that in the 2D Ising
model, because the critical divergence follows a power law in the
former while it follows a logarithmic law in the latter. However,
the partial entropy method still works very well in this quantum
system.  The Heisenberg Hamiltonian reads, $H=- 2\sum_{\langle ij
\rangle } {\vec S}_{i} \cdot {\vec S}_{j}$, where ${\vec
S}_i={\vec\sigma}_i/2$ is the spin-1/2 operator and $\langle ij
\rangle$ indicates bonds between nearest neighbor sites.  This
model has a second-order phase transition at the critical
temperature $T_c = 1.6778 \pm 0.0002$ as determined by
high-accuracy quantum Mote Carlo simulation and phenomenological
renormalization-group analysis \cite{souza}.

The reduced density matrix for a nearest neighbor pair of sites
can again be found in a simple form, $\rho_p(T)=1/4-[2u(T)/9]{\vec
S}_{1} \cdot {\vec S}_{2}$, based on a symmetry analysis
\cite{corr, hai}.  Its coefficient is related to the internal
energy density $u(T)$ by tracing the reduced density matrix with
the Hamiltonian.  The partial entropy can then be expressed as
$S_p(T)=- \{[9-2u][\ln(9-2u)-2\ln6] + [3+2][\ln(3+2)-
\ln12]\}/12$.  These expressions remain valid for finite system
sizes under periodic boundary conditions. We use quantum Mote
Carlo simulation with a stochastic series expansion algorithm
\cite{mon} to calculate the density of internal energy and then to
obtain the partial entropy from the above formula.

We will show that this critical temperature can be obtained
accurately by using the partial entropy method with the
calculation for small-size system.

\begin{figure}[ht]
\begin{center}
\includegraphics[height=6cm,width=8cm]{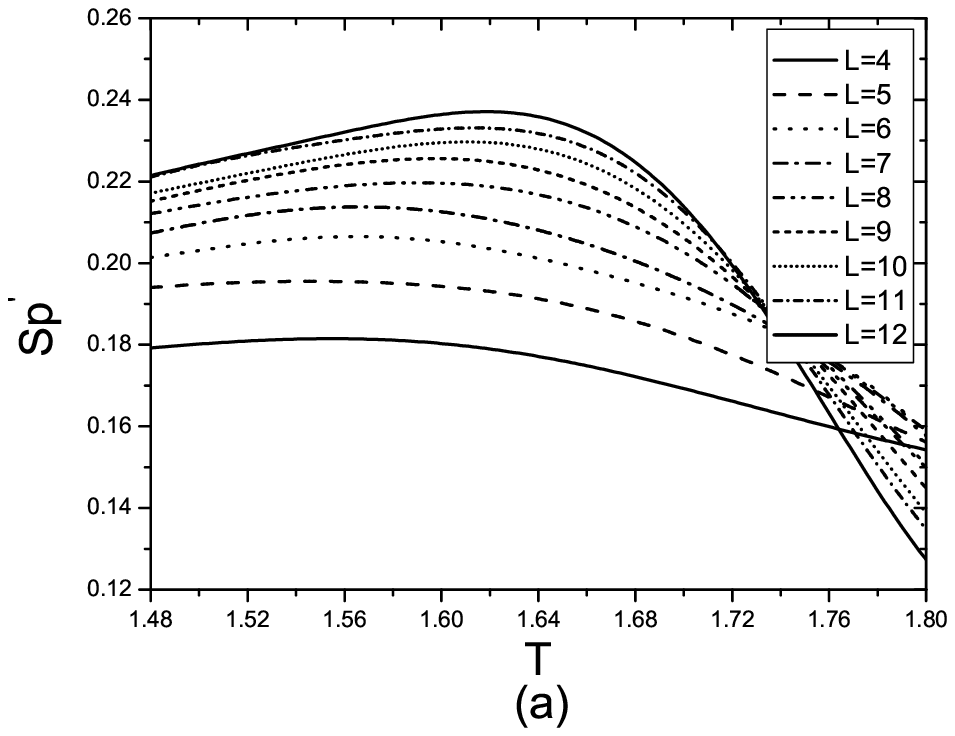}\\
\includegraphics[height=6cm,width=8cm]{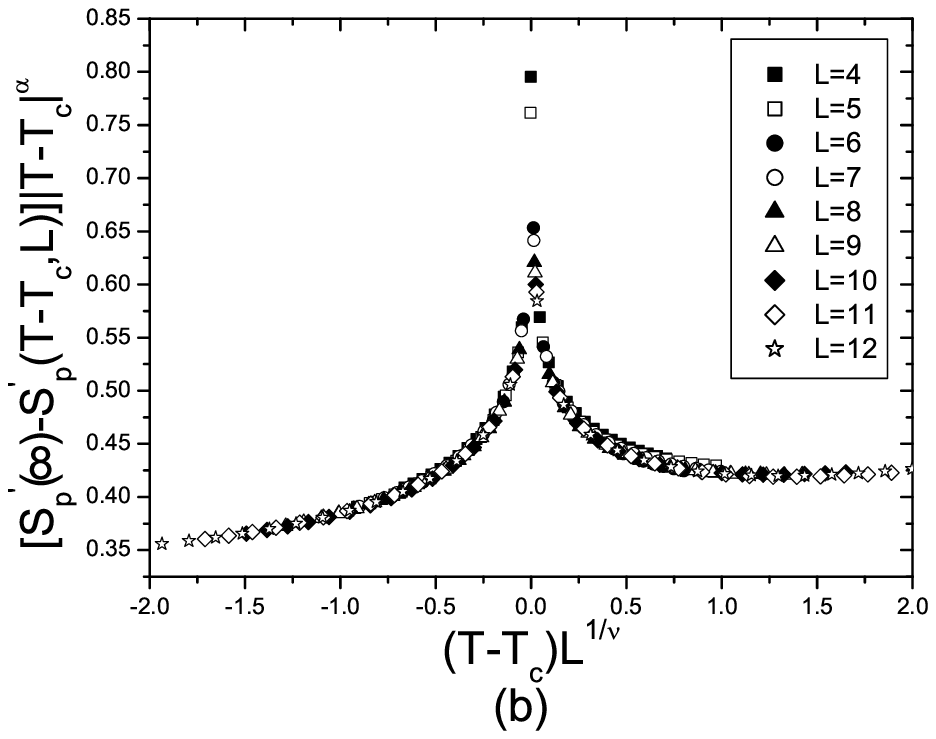}
\end{center}
\caption{(a) The derivative of partial entropy shows a peak at
certain temperature. (b) The finite-size scaling of the derivative
of partial entropy.} \label{fig5}
\end{figure}

As before, the partial entropy is monotonically increasing with
$T$, while its derivative $S^{\prime}_{p}(T)$ arrives at a maximum
at certain temperature.  The peaks of $S^{\prime}_{p}(T)$ for
linear sizes of $L=4-12$ are shown in detail in Fig.
\ref{fig5}(a).  The peaks sharpen with increasing $L$, and suppose
to become singular at $L\rightarrow \infty$.  The different curves
can be collapsed on to a single one, shown in Fig. \ref{fig5}(b),
by the following scaling relation
\begin{eqnarray}
&& S^{\prime}_p(\infty)=S^{\prime}_p(L)+g L^{\alpha/\nu},
\nonumber \\
&& [S^{\prime}_p(t,L)-S^{\prime}_p(\infty)]|t|^{\alpha}= f( t
L^{1/\nu}),
\end{eqnarray}
where $t=1-T/T_c$, $f$ is an universal function and $g$ is a
constant.  Our result yields the critical exponents
$\alpha=-0.1116 \pm 0.0005$, $\nu=0.705 \pm 0.003$ and the
critical temperature $T_c=1.677 \pm 0.001$, which agree very well
with those obtained earlier \cite{le,souza}.

More remarkably, the finite-size scaling for the partial entropy
starts to work from relatively small system sizes.  If we use data
only from $L=4-8$, the fitting to a scaling function then yields
$\alpha=-0.703\pm 0.003$, $\nu=0.1196 \pm 0.0005$ and $Tc=1.678
\pm 0.002$, which are already very good.  On the other hand,
thermodynamic quantities, such as the specific heat, show good
scaling behavior only for $L \geq 8$ \cite{sandv}.

\par In conclusion, we suggest that the partial entropy is quite an
effective tool to quantify the finite-temperature phase
transitions. The critical temperature can be determined by the
inflexion of the partial entropy or the peak of its derivative
with respect to temperature.  The partial entropy shows good
finite-size scaling even for relatively small system sizes. The
key point lies in the fact that the partial entropy catches the
most essential feature of phase transitions, i.e., the information
of critical fluctuations. The advantages of this method are: one
does not need a pre-assumed order parameter; does not need to
study a special large-scale correlation function, and it is
straightforward and efficient. Elements of the partial entropy are
closely related to the thermodynamic quantities or correlation
functions via simple symmetry analysis.

J. Cao and Y. Wang are supported by NSFC under grant No.10474125,
No.10574150 and the national key project on basic research. Q. Niu
is supported by the NSF and the R.A. Welch Foundation.

$^*$Electronic address: yupeng@aphy.iphy.ac.cn

\end{document}